\documentclass[doublecol]{epl2} 
\usepackage{psfrag,epsfig,amsfonts,amssymb,amsmath,wasysym}
\usepackage{dcolumn}

\usepackage[normalem]{ulem}
\usepackage{color}

\title{Anisotropic diffusion in square lattice potentials: 
giant enhancement and control}
\shorttitle{Anisotropic diffusion in square lattice potentials} 

\author{David Speer\inst{1} \and Ralf Eichhorn\inst{2} \and Peter Reimann\inst{1}}
\shortauthor{D. Speer \etal}

\institute{                    
  \inst{1} Universit\"at Bielefeld, 
Fakult\"at f\"ur Physik, 33615 Bielefeld, Germany\\
  \inst{2} NORDITA, Roslagstullsbacken 23, 10691 Stockholm, Sweden
}

\pacs{05.45.-a}{Nonlinear Dynamics and nonlinear dynamical systems}
\pacs{05.40.-a}{Fluctuation phenomena, random processes, noise, and Brownian motion}
\pacs{05.60.-k}{Transport processes}

\abstract{
The unbiased thermal diffusion of an overdamped Brownian 
particle in a square lattice potential is considered
in the presence of an externally applied ac driving.
The resulting diffusion matrix exhibits two orthogonal 
eigenvectors with eigenvalues $D_1>D_2>0$,
indicating anisotropic diffusion along a 
``fast'' and a ``slow principal axis''.
For sufficiently small temperatures, 
$D_1$ may become arbitrarily large and 
at the same time $D_2$ arbitrarily small.
The principal diffusion axis can be made to 
point into (almost) any direction
by varying either the driving amplitude
or the coupling of the particle to the potential,
without changing any other 
property of the system or the driving.
}

\begin{document} 

\maketitle

\section{Introduction}
Thermal diffusion plays a key role in many
physical, chemical, and biological processes.
At thermal equilibrium, the free diffusion
as considered, e.g., by Einstein \cite{ein05},
is generically reduced by an additional periodic 
potential \cite{lif62}, and -- at least for 
the most common periodic lattice structures --
remains spatially isotropic (see below).
Out of equilibrium,
the main focus has so far been on periodic 
potentials in one-dimension, perturbed by either
a static ``bias force'' (tilted washboard potentials) 
\cite{tilt,lub99,lin08}
or by an unbiased, time-dependent driving
\cite{ac}.
The most remarkable finding in both cases is 
that the diffusion coefficient $D$ may
exhibit a giant enhancement over the
free (Einsteinian) diffusion coefficient $D_0$.
More precisely, for asymptotically small $D_0$,
the ratio $D/D_0$ diverges, 
while $D$ itself still tends to zero,
apart from certain fine-tuned (non-analytical)
potential shapes and system parameters, 
for which $D$ may remain finite.

Another topic of considerable recent interest 
concerns the effect of disorder 
(random deviations from a strictly periodic potential),
which may give rise to a further diffusion enhancement
or even to anomalous diffusion \cite{disorder}.

Works on genuine new phenomena in higher dimensions
are scarce:
Guantes and Miret-Art\'es \cite{gua03} studied 
biased diffusion induced by an asymmetric external 
driving.
Sancho et al. and Lacasta et al. \cite{san04} put 
their main focus on (possibly transient) 
anomalous diffusion effects.
Experimentally, Tierno et al. observed 
giant transversal diffusion 
of paramagnetic particles 
on an uniaxial garnet film 
in the presence of an oscillating magnetic field 
\cite{tie10}.

Here, we reconsider the prototypical situation 
of Brownian motion in a more than one-dimensional
periodic potential and we demonstrate that a 
simple ac driving generically results in
anisotropic diffusion, whose direction and 
magnitude depend in a quite intriguing way 
on the potential, the driving, and the particle 
properties.

\section{Model}
To keep things as simple as possible, we specifically
focus on a square lattice in two-dimensions and 
on a 
square-wave (rectangular) driving.
Yet, our main results readily carry over to
various other drivings and
more general lattice potentials in two as well 
as in three dimensions.

Our starting point is the the Langevin dynamics
\begin{equation}
M \ddot{\vec r}(t)+\eta \dot{\vec r}(t) = 
-\vec{\nabla} U(\vec r(t)) +\vec{F}(t) 
+ \sqrt{2\eta kT} \vec\xi (t) \ ,
\label{1}
\end{equation}
where $M$ is the mass of the Brownian particle,
$\eta$ its friction coefficient,
and $\vec r= x \vec e_x+ y\vec e_y$ its 
position in the $x$-$y$-plane,
with $\vec e_x$ and $\vec e_y$ being the
unit vectors in $x$- and $y$-direction.
Further, $U(\vec r)$ stands for the spatially periodic potential,
$\vec{\nabla} U(\vec r)$ its gradient, and
$\vec F(t)$ the temporally periodic force.
The last term in (\ref{1}) accounts for 
thermal noise as usual \cite{lif62,tilt,lub99,ac,gua03,san04}:
$T$ is the temperature, $k$ Boltzmann's constant,
and $\vec \xi(t)=\xi_x(t)\vec e_x+\xi_y(t)\vec e_y$ 
consists of two independent, delta-correlated, 
Gaussian noises $\xi_x(t)$, $\xi_y(t)$.

\begin{figure}
\epsfxsize=0.7\columnwidth
\epsfbox{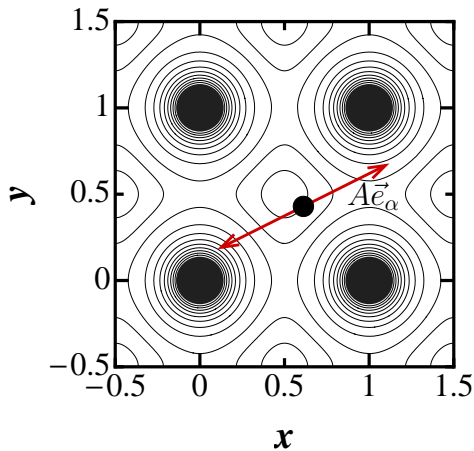} 
\caption{
Schematic illustration of the periodic potential 
from (\ref{2}), (\ref{3}) (black contour lines) 
and of the square-wave driving from (\ref{4}) (red double-arrow)
which act on the Brownian particle according 
to (\ref{1}) (black dot).
The depicted choice of parameters $L=1$, $Q=1$, 
and $\alpha=35^\circ$ represents our ``standard
example'' in eqs. (\ref{5}), (\ref{6}) and in
figs.\ \ref{fig2} - \ref{fig7}.
One of the periodic potential minima is located at the 
center of the plot.
}
\label{fig1}
\end{figure}

As announced, we specifically address the case
that the periodic potential 
$U(\vec r)$ is composed of rotation-symmetric
on-site potentials $U_1(\vec r)$ on a square 
lattice with period $L$,
\begin{equation}
U(\vec r)=\sum_{m,n=-\infty}^\infty 
U_1(\vec r -[m\vec e_x+n\vec e_y]L) \ .
\label{2}
\end{equation}
As illustrated by fig.\ \ref{fig1},
our standard example will be the Yukawa potential
\begin{equation}
U_1(\vec r)= Q\ \exp\{-|\vec r|/4L\}/|\vec r| \ ,
\label{3}
\end{equation}
where $Q\geq 0$ quantifies the coupling of the particle to the 
potential (e.g. the particle charge in the case of a
screened electrostatic potential).

Likewise, the announced square-wave driving is formally given by
\begin{equation}
\vec F(t)=\vec e_\alpha \, A \ \mbox{sign}[\sin (\omega t)] 
\label{4}
\end{equation}
with amplitude $A$, frequency $\omega$,
and direction $\vec e_\alpha:=\vec e_x\cos\alpha+\vec e_y\sin\alpha$.
In other words, it switches between $A$ and $-A$ every 
half-period $\pi/\omega$, and $\alpha$ represents the angle 
between the driving force and the $x$-axis, 
see fig.\ \ref{fig1}.

Yet, as already said, our main qualitative findings turn out 
to be largely independent of the above particular choice of 
on-site potential, lattice structure, and driving.

To further reduce the number of model parameters, 
we can and will choose the units of length, time, 
energy, and temperature so that
\begin{equation}
L=1\ ,\ \ \omega=1\ ,\ \ \eta=1 \ ,\ \ k=1 \ ,
\label{5}
\end{equation}
and we focus on the simplest and most important case 
\cite{lif62,tilt,lub99,ac} 
that inertia effects are negligible
(overdamped limit), i.e. $M\to 0$ 
in (\ref{1}), yielding 
\begin{equation}
\dot{\vec r}(t) = 
-\vec{\nabla} U(\vec r(t)) +\vec{F}(t) 
+ \sqrt{2  T} \vec\xi (t) \ .
\label{6}
\end{equation}

We remark that in the presence of an
additional, externally applied static 
bias force, a quite interesting response 
behavior in the form of a directed net 
particle motion arises \cite{spe09,spe11}.
In the following, our main focus will 
be on a quite different issue, namely 
the unbiased diffusion of a Brownian 
particle in the absence of any further 
perturbation.

\section{Diffusion matrix}
For symmetry reasons, the above specified dynamics rules out any 
systematic particle transport, i.e we are dealing with
a purely diffusive, unbiased Brownian motion,
characterized by a $2\times 2$ diffusion matrix ${\bf D}$
with matrix elements
\begin{equation}
D_{ij} = \lim_{t\rightarrow\infty} 
\frac{\langle r_i(t)\,r_j(t)\rangle}{2\, t} \ ,
\label{7}
\end{equation}
where $r_i,\,r_j\in\{x,\, y\}$ 
(see below (\ref{1}))
and where $\langle\cdot\rangle$ indicates the ensemble average
over many realizations of the stochastic dynamics.
For ergodicity reasons and due to the long-time limit in (\ref{7}), 
the initial conditions are irrelevant and we can
without loss of generality focus on
\begin{equation}
\vec r(0)=\vec 0 \ .
\label{8}
\end{equation}

\begin{figure}
\epsfxsize=0.8\columnwidth
\epsfbox{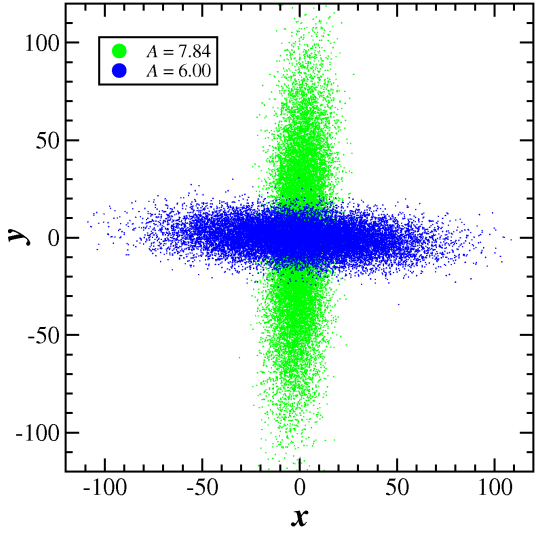} 
\caption{
Typical examples of diffusive ``particle clouds''.
The corresponding (unbiased) Gaussian probability
densities are completely characterized by the 
variance $2D_1t$ along the long principal axis and 
by the variance $2D_2t$ along the short principal axis
of the ellipsoidal clouds. In other words, the
plot must simply be rescaled by $\sqrt{t}$
in the course of time.
Quantitatively, every single depicted point represents
a numerical solution at time $t=10^5$ of the stochastic 
dynamics (\ref{2})-(\ref{6}) with initial 
condition (\ref{8}) and parameters $Q=1$, $\alpha=35^\circ$,
$T=5\cdot 10^{-5}$, and $A=6$ (blue), $A=7.84$ (green).
}
\label{fig2}
\end{figure}

Since ${\bf D}$ is symmetric and positive, there are two 
orthogonal eigenvectors with eigenvalues 
\begin{equation}
D_1 \geq D_2>0 \ .
\label{9}
\end{equation}
Essentially, a statistical 
ensemble of (non-interacting) 
Brownian particles, all starting from the origin at time $t=0$,
thus evolves in the course of time into a bigger 
and bigger ``particle cloud'' of ellipsoidal shape, 
see fig.\ \ref{fig2}.
For large times $t$, and neglecting local details on 
the scale of the spatial period $L$, this cloud is 
quantified by a Gaussian probability density \cite{lub99,ris84}
of variance $2D_1t$ along some ``fast principal axis'' 
$\vec e_\Theta:=\vec e_x\cos\Theta+\vec e_y\sin\Theta$,
and of variance $2D_2t$ along the orthogonal ``slow principal axis''.
The angle $\Theta$ between ``fast direction'' and $x$-axis
is thus $180^\circ$-periodic, i.e.
\begin{equation}
\Theta \ \mbox{and}\  \Theta+180^\circ \ \mbox{are equivalent.}
\label{10}
\end{equation}

\begin{figure}
\epsfxsize=0.9\columnwidth
\onefigure{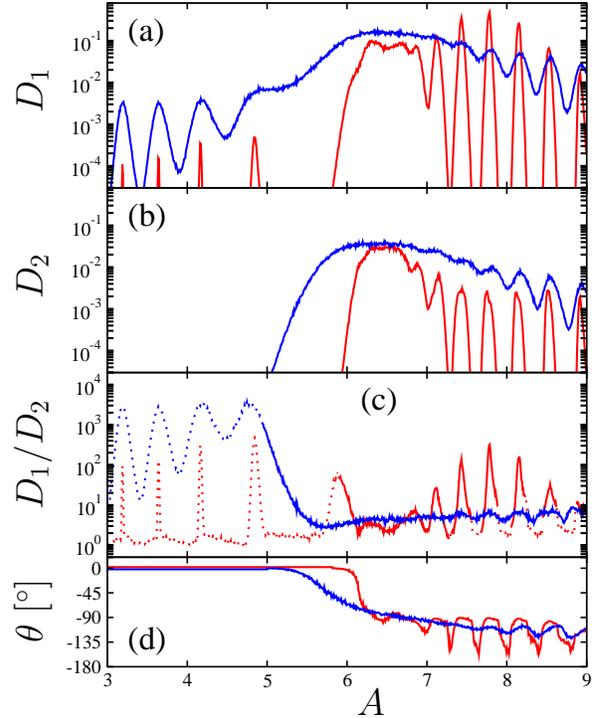}
\caption{
Eigenvalues $D_{1,2}$ of the diffusion matrix (\ref{7})
and angle $\Theta$ between first eigenvector and 
$x$-axis (see also (\ref{9},\ref{10}))
versus driving amplitude $A$,
obtained  
 -- similarly as in fig. \ref{fig2} --
from 800 numerical realizations 
over $10^5$ time units of the stochastic dynamics
(\ref{2})-(\ref{6}) with initial condition (\ref{8})
and parameters $Q=1$, $\alpha=35^\circ$, and  
$T=5\cdot 10^{-4}$ (blue), $T=5\cdot 10^{-5}$ (red).
In spite of the very large final time $t=10^5$,
the average particle displacement along the ``slow principal axis''
(and possibly also along the ``fast'' one)
was less than one spatial period $L=1$ for some 
$A$-values.
The corresponding $D_2$ (and $D_1$) values are below 
the range depicted in (b) and (a).
In (c), the corresponding results are marked 
by dotting the curves.
They do not represent the ``true'' limits 
in (\ref{7}), since $t=10^5$ is still too 
small. On the other hand, sufficiently large
$t$-values are way beyond our numerical capabilities.
In (d) we did not dot the corresponding $A$ regions since
we are quite confident that the depicted results are
already very close to the ``true'' long time limit.
}
\label{fig3}
\end{figure}

Free diffusion (vanishing $U(\vec r)$ and $\vec F(t)$)
according to Einstein \cite{ein05} 
amounts to $D_1=D_2=D_0$ with $D_0=kT/\eta$ for the 
original dynamics (\ref{1}) and $D_0=T$ for the 
rescaled dynamics (\ref{6}).

Including the periodic potential $U(\vec r)$,
but still without oscillating force $\vec F(t)$, 
the system (\ref{2})-(\ref{6})
is still invariant under rotations by $90^\circ$, 
implying isotropic diffusion ($D_1=D_2$).

Including also the driving $\vec F(t)$, 
this symmetry is broken, implying anisotropic 
diffusion ($D_1>D_2$) in the generic case.
The remaining two symmetry arguments are:
(i) It is sufficient to consider 
$\alpha\in[0^\circ,\, 45^\circ]$.
(ii) If the driving direction $\vec e_\alpha$ 
coincides with a symmetry axis of the square 
lattice (i.e. $\alpha=0^\circ$ or $\alpha=45^\circ$)
then $\vec e_\Theta$ will be parallel or 
orthogonal to $\vec e_\alpha$.
In any other case ($0^\circ<\alpha<45^\circ$)
it is impossible to predict $\vec e_\Theta$ a priori.

After a couple of unsuccessful attempts one 
furthermore realizes that analytical progress is quite 
hopeless.
Therefore we now turn to the presentation of our
numerical results (next Section), followed by a brief account
of the underlying basic physical mechanisms (Section `Discussion').

\section{Results}
We first exemplify the most interesting 
regime of the parameters $Q$, $A$, $\alpha$
and later provide a more systematic exploration:
fig. \ref{fig3} depicts the ``fast'' and ``slow''
diffusion coefficients $D_1$ and $D_2$ together
with their ratio $D_1/D_2$ and the 
``fast direction''  $\Theta$.
A pronounced anisotropy of the diffusion is 
indicated by  the fact that 
$D_1\gg D_2$.
The corresponding  ``particle cloud'' 
(see fig.\ \ref{fig2} and below eq. (\ref{9}))
thus actually resembles a very narrow ``particle beam''.
Its orientation $\Theta$ may assume any value between
$0^\circ$ and 
(almost) $-180^\circ$
upon variation of the driving amplitude $A$
for the specific choice of the remaining parameters used in fig. \ref{fig3}.
For other such choices, an even larger range of $\Theta$-values can be covered.
Since $\Theta$ and $\Theta+180^\circ$ are equivalent (see (\ref{10})),
we can conclude that essentially any possible orientation 
of the ``fast'' diffusion direction can be realized
upon variation of $A$, without changing any other 
property of the system or the driving, 
see fig. \ref{fig2}.

\begin{figure}
\epsfxsize=1.0\columnwidth
\epsfbox{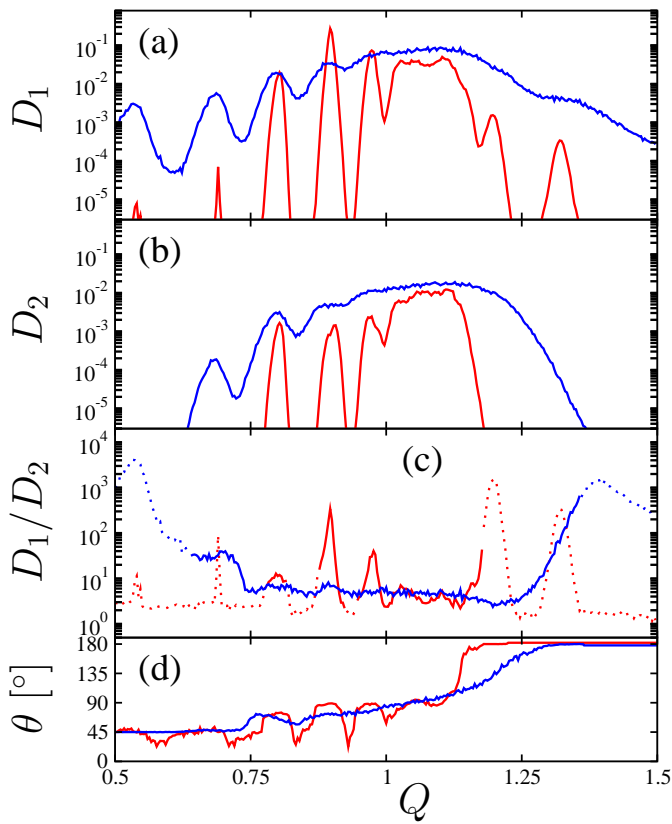} 
\caption{
Same as fig. \ref{fig3} but for varying coupling
strength $Q$ and fixed driving amplitude $A=7$.
}
\label{fig4}
\end{figure}

Decreasing the temperature $T$ yields larger
$D_1/D_2$ ratios for practically 
all amplitudes $A$ in fig. \ref{fig3} 
for which the long time limit in (\ref{7}) could be numerically reached,
i.e. the diffusion anisotropy gets more pronounced.
Whereas $D_2$ always decreases upon lowering the temperature,
$D_1$ increases within certain ``$A$-windows''
around $A=7.4$, $A=7.8$, $A=8.2$ etc., and decreases otherwise.

According to fig. \ref{fig4}, analogous
findings are recovered upon variation of the 
potential strength $Q$ from (\ref{3}).
Hence, particles with different ``charges''
$Q$ can be diffusively ``beamed'' 
into different directions very ``fast'' and
with very high precision.
In other words, a periodic ``surface potential'' 
can act as a very efficient particle sorting
device.

\begin{figure}
\epsfxsize=1.0\columnwidth
\epsfbox{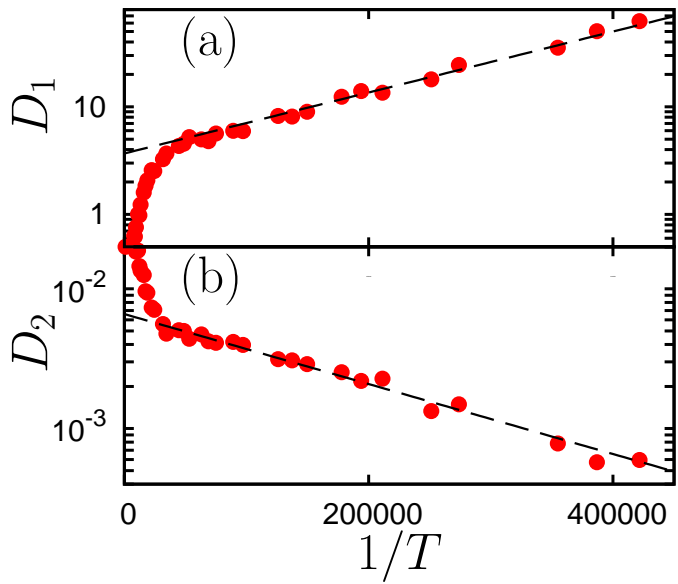} 
\caption{
Eigenvalues $D_{1,2}$ of the diffusion matrix (\ref{7}),
presented as Arrhenius-plot ($\log D_i$ versus $1/$temperature),
from 1000 numerical realizations 
over $10^5$ time units of the stochastic dynamics
(\ref{2})-(\ref{6}) with initial condition (\ref{8})
and parameters $Q=1$, $\alpha=35^\circ$, and $A=7.77$.
Dashed: Asymptotic eq. (\ref{11}).
}
\label{fig5}
\end{figure}

A more detailed low-temperature asymptotics 
of $D_{i}$ ($i=1,\,2$) is provided by fig. \ref{fig5}, 
evidencing an Arrhenius-type behavior
\begin{equation}
D_i\sim \exp\{- E_i/T\}
\label{11}
\end{equation}
with certain ``effective energies'' $E_1<0< E_2$.
The same behavior is recovered within all of 
the above mentioned ``$A$-windows''.
For all other $A$-values,
one finds (as expected and hence not shown)
the same asymptotics as in (\ref{11}), 
but now with $0<E_1<E_2$.

We thus can conclude that the diffusive
particle motion can become arbitrarily ``fast'' 
and anisotropic for sufficiently small 
temperatures $T$.
Likewise, in comparison to the free diffusion coefficient 
$D_0=T$, the ratio $D_1/D_0$ diverges and $D_2/D_0$
approaches zero for asymptotically small $T$.

\begin{figure}
\epsfxsize=0.7\columnwidth
\epsfbox{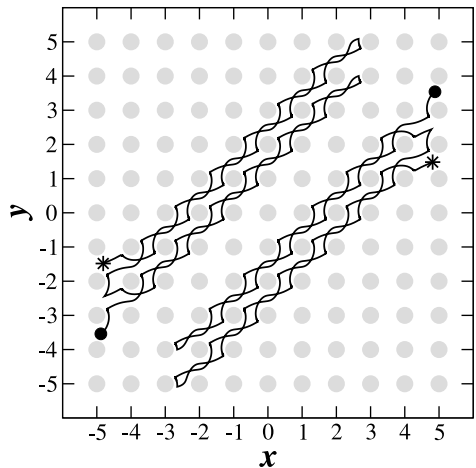} 
\caption{
A pair of transporting attractors, arising due to 
spontaneous symmetry breaking.
Each of the two black trajectories represents a solution 
of the deterministic dynamics (\ref{2})-(\ref{6}) with 
$Q=1$, $A=7$, $\alpha=35^\circ$, and $T=0$.
The ``star'' of the lower right trajectory 
indicates the initial position
at $t=\pi$ and the ``dot'' the final positions at 
$t=5\,\pi$, i.e. after 2 temporal periods of the 
square-wave driving (\ref{4}). 
At the end of the depiced evolution, the solution
has advanced along the $y$ axis by exacly two spatial
periods of the periodic potential 
(indicated by the grey dots, see also fig. \ref{fig1}).
In other words, we are dealing with a periodic 
attractor with $v=1$ and $\phi=90^\circ$.
Likewise, the upper left trajectory advances between
$t=0$ and $t=4\, \pi$ into the opposite direction.
}
\label{fig6}
\end{figure}

\begin{figure}
\epsfxsize=1.0\columnwidth
\epsfbox{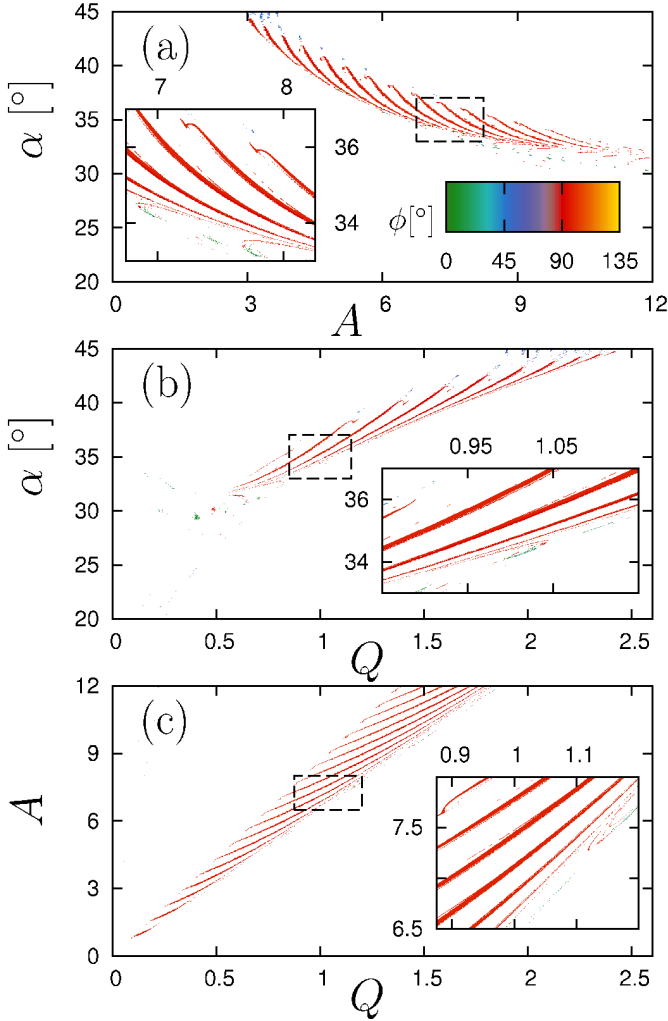} 
\caption{
Deterministic attractors by solving (\ref{2})-(\ref{6}) with $T=0$ 
numerically for a large number of randomly sampled initial conditions
$\vec r(0)$.
(a) Varying driving angle $\alpha$ and amplitude $A$
at fixed coupling strength $Q=1$.
(b) Varying $\alpha$ and $Q$ at fixed $A=7$.
(c) Varying $A$ and $Q$ at fixed $\alpha=35^\circ$.
White: All numerically detected attractors
are non-transporting
(vanishing net velocity $\vec v=\lim_{t\to\infty}\vec r(t)/t$).
Colored: There is a symmetric pair of attractors 
exhibiting spontaneous transport of 
opposite velocities 
(cf. fig. \ref{fig6}).
The angle $\phi$ between these velocities 
and the $x$-axis is indicated by the different colors.
The overwhelming majority of the colored pixels
corresponds to $\phi=90^\circ$ (red), but also
a few blue and green pixels are still 
discernible.
In most cases, the modulus of the velocities 
$|\vec v|$ 
was found to be of order one (not shown).
Insets: Magnification of dashed rectangles.
}
\label{fig7}
\end{figure}

\section{Discussion}
For an intuitive understanding of what is going on, 
we first focus on the dynamics 
(\ref{2})-(\ref{6}) in the deterministic limit ($T=0$).
Without driving ($A=0$) the particle readily settles down
into one of the local minima of the periodic potential
(\ref{2}), (\ref{3}), 
see also fig.\ \ref{fig1}.
Upon increasing $A$, all those point attractors 
evolve via a complicated sequence of bifurcations 
into (possibly coexisting)
periodic, quasi-periodic, or chaotic attractors, 
whose quite intricate details also depend on the 
values of $Q$ and $\alpha$ \cite{spe09,spe11}.
As might not have been immediately anticipated,
but in fact is rather plausible at second glance,
some of the so emerging deterministic attractors 
give rise to a permanent directed particle motion,
see fig. \ref{fig6}.
For symmetry reasons, such ``transporting attractors'' 
always appear in pairs with opposite transport 
velocities by way of spontaneous symmetry breaking.
Besides such a pair, there may or may 
not coexist additional (transporting or 
non-transporting) attractors.
The actual occurrence of those transporting attractors and 
the concomitant quantitative transport velocities 
$\vec v$
are only accessible by numerical means:
fig. \ref{fig7} depicts such numerical results for three 
representative cross sections through the full, 3-dimensional 
$\alpha$-$A$-$Q$ parameter space.

Turning to small but finite temperatures $T$, the above
deterministic attractors become metastable states since the
thermal noise now induces rare transitions between them.
For weak to moderate driving amplitudes $A$, these 
are mainly ``hopping'' transitions between 
neighboring potential wells in $x$- and 
$y$-direction 
(cf. fig.\ \ref{fig1})
at certain rates $\gamma_x$ and $\gamma_y$, respectively.
Since the driving $\vec F(t)$ predominantly
acts along the $x$-direction for
$0<\alpha<45^\circ$, it seems plausible 
that $\gamma_x>\gamma_y$.
Hence one can infer that
\begin{eqnarray}
D_1 & \approx & L^2\gamma_x/2\ ,
\label{12}
\\
D_2 & \approx & L^2\gamma_y/2 \ .
\label{13}
\end{eqnarray}
Therefore, we can conclude that 
$\Theta\approx 0^\circ$.
From the usual Boltzmann-Arrhenius
temperature dependence of the rates 
$\gamma_{x,y}$ one finally recovers (\ref{11}) 
with $0<E_1<E_2$.
For the rest, the rates $\gamma_{x,y}$ and
hence $D_{1,2}$ from (\ref{12},\ref{13})
may still depend on $A$ 
(and likewise on $Q$ and $\alpha$)
in a very complicated way.
Altogether, these considerations
explain the main features of fig. 1 
up to about $A=6$.

Next we focus on the ``red stripes'' in fig. 
\ref{fig7}, indicating two symmetric
attractors with spontaneous transport in 
$y$-direction at certain velocities $\pm v$.
Denoting by $\gamma_y$ the rate of 
noise induced transitions between them,
a similar calculation as in Ref. \cite{lin08,kap82} 
yields for the diffusion coefficient in 
$y$-direction the approximation
\begin{equation}
D_1\approx v^2/(2\gamma_y) \ ,
\label{14}
\end{equation}
whereas in $x$-direction one obtains similarly as
in (\ref{12}) the approximation
\begin{eqnarray}
D_2 & \approx & L^2\gamma_x/2 \ .
\label{15}
\end{eqnarray}
Hence, one can conclude that $\Theta \approx \pm 90^\circ$.
This provides an explanation 
of eq. (\ref{11}) with $E_1<0<E_2$, 
an explanation of fig. \ref{fig3} within 
the above mentioned ``$A$-windows'', 
and an explanation of fig. \ref{fig5}.
Moreover, we recognize that these ``$A$-windows'' 
are in fact the descendants of the ``red stripes''
in fig. \ref{fig7}.

So far, we tacitly neglected the 
possibility that besides the transporting 
attractors there may still coexist further
non-transporting attractors.
Numerically, one finds that this is 
indeed the case within a subset of the 
colored regions in fig. \ref{fig7}
(not shown).
In such a case, the noise induced transitions between the 
various attractors are governed by several 
different rates, resulting in similar but more 
complicated estimates for $D_{1,2}$ than in (\ref{12})-(\ref{15}).

Analogous generalizations apply for the small
parameter regions in fig. \ref{fig7} which
are colored differently from red, corresponding 
to a spontaneous deterministic transport into a 
direction substantially different from 
$\phi= 90^\circ$.

Coming back to the remaining $A$-values in fig. \ref{fig3}
-- namely those larger than about $A=6$ but not contained
in one of the $A$-windows --
it turns out that, essentially, again thermal 
hopping prevails, but now predominantly
in steps of $\sqrt{2}L$
along the bisectrix $y=x$,
hence $\Theta\approx -135^\circ$.
Occasionally, one also encounters 
more complicated types of motion, 
resulting in even smaller 
$\Theta$-values.

As expected, with increasing temperature $T$,
all these feature become more and more ``washed
out'' in fig. \ref{fig3}. Especially, this 
applies to the neighborhood of $A=6.5$,
in accordance with the tightly nested 
red and white stripes in the corresponding 
region of fig. \ref{fig7}.

Analogous considerations apply to fig. \ref{fig4}.
Overall, we thus can conclude that for sufficiently low 
temperatures $T$, giant anisotropic diffusion arises within 
all colored parameter regions in fig. \ref{fig7}, 
and that pronounced variations of the principal diffusion 
axes are expected whenever crossing a border between 
colored and white regions.

The entire pertinent parameter regions in fig. \ref{fig7} 
are not very large as far as their measure is concerned,
but still very appreciable for instance from the
following viewpoint:
For (almost) any given amplitude $A$
it is possible to find angles $\alpha$ 
and coupling strengths $Q$
for which all those effect can be 
observed \cite{spe11}.

Finally, it also seems worth mentioning that
in contrast to giant diffusion enhancement 
in one-dimensional periodic potentials \cite{tilt,ac},
in our present case (i) not only the ratio $D_1/D_0$, 
but also $D_1$ itself diverges for asymptotically 
small temperatures, and (ii) the pertinent parameter 
regions are of finite measure.

\section{Conclusions}
Starting with Einstein's ground breaking work on Brownian 
motion \cite{ein05}, the subject of diffusion has 
attracted a lot of attention due to its practical 
relevance in numerous specific systems, but 
also due to its theoretical interest as a fundamental 
form of transport per se. 
Here, we reconsidered diffusion in a particularly simple
``minimal model'': a spatially periodic system, 
driven out of equilibrium in one of the experimentally 
most natural and common ways, namely by an oscillating 
driving force.
Our first main finding is an extreme anisotropy of the diffusion, 
whose ``speed'' in one direction may grow exponentially 
fast as temperature decreases, while the perpendicular 
``speed'' approaches zero exponentially fast. 
Our second main finding is that the 
principal axes of the anisotropic diffusion can be made
to point into (practically) any direction by 
solely varying e.g. the ac driving amplitude
or the coupling of the particle to the 
periodic potential,
but without changing any other 
property of the system or the driving.

The indispensable prerequisites for those
diffusion effects are
very weak and ubiquitous: 
Brownian motion in a spatially
periodic lattice potential under the 
action of an oscillating driving 
force is a paradigmatic scenario
in a large variety of different contexts,
e.g. diffusion on the surface \cite{gua03}
but also in the bulk of a crystal
under the action of electromagnetic
perturbations,
migration in periodic arrays of laser
traps, or in periodically structured
micro- or nano-fluidic systems, 
to name but a few \cite{san04,tie10,spe11}.

One reasons why those very fundamental diffusion
phenomena have not been discovered earlier may be the fact
that the underlying key mechanisms of 
spontaneous symmetry breaking has not been
naturally associated with such systems up to now.
A second reason is that a very careful
numerical exploration is required with
a clear idea of what one is actually 
looking for.
In turn, once the quantitative numerical 
predictions for a specific experimental 
system are available, verifying and 
exploiting the predicted effects seems 
quite feasible and worth while to us.

\acknowledgments

This work was supported by Deutsche Forschungsgemeinschaft
under SFB 613 and RE1344/5-1

\end{document}